\documentclass[prb,twocolumn,showpacs]{revtex4}
\usepackage{graphicx}
\usepackage{amsfonts}
\usepackage{amsmath}
\usepackage{amssymb}
\usepackage{bm}
\usepackage{hyperref}
\usepackage{slashed}
\begin{document}
\title{Spinon Majorana fermions}
\author{Rui Wang$^{1,2}$, Hong-Yan Lu$^{1,3}$, Baigeng Wang$^2$, C. S. Ting$^1$}
%\author{Baigeng Wang$^{2}$}
%\author{L. Sheng$^{2}$}
%\author{D. Y. Xing$^{2}$}
%\author{Jian Wang$^{1,3}$}
\affiliation{$^1$ Department of Physics and Texas Center for Superconductivity, University of Houston, Houston, Texas 77204, USA\\
$^2$ National Laboratory of Solid State Microstructures, Collaborative Innovation Center of Advanced Microstructures, and Department of Physics, Nanjing University, Nanjing 210093, China\\
$^3$ School of Physics and Electronic Information, Huaibei Normal University, Huaibei 235000, China
}
\date{\today }

\begin{abstract}
A realization of Majorana fermions is proposed in the frustrated magnets via the topological proximity effect. Specifically, we consider a theoretical model, where a topological insulator is coupled to a frustrated magnetic material through the spin exchange interaction.  Using the renormalization group-based self-consistent mean-field approach, and calculating the self-energy correction due to the topological insulator, we find that the spin texture and the spin-momentum locking of the Dirac cone will be inherited by the spinons in the nearby frustrated magnets. This leads to a particular topological state of matter that supports the Majorana excitations. Unlike the conventional realization in SC systems, these Majorana fermions are the combination of spinons and anti-spinons, rather than electrons and holes. They can participate in the transport of spinons, leading to nontrivial properties of the spin transport.
\end{abstract}

\maketitle

\section{introduction}
Majorana fermions (MFs), which are their own anti-particles, have attracted great interest in condensed matter physics due to their promising application in the fault-tolerant quantum computation \cite{Kitaev,Nayak}. It is natural to describe the MFs using a fermion operator $\gamma$ that is related to the electron operators $\psi$ by $\gamma=(\psi^{\dagger}+\psi)/2$, so that the $\gamma^{\dagger}=\gamma$. Since the Bogoliubov quasi-particles of superconductor (SC) are combination states of electrons and holes, SC constitutes the best platform to realize MFs. Theoretically, many proposals have been put forward: Read and Green investigated the $p_x\pm ip_y$ superconductor \cite{Read}, where the Majorana bound state was suggested to exist in a vortex defect. Fu and Kane \cite{Fu} showed that $p_x\pm ip_y$-wave-like pairing may occur in the surface of a strong topological insulator (TI) when an ordinary s-wave superconductor is nearby. Sau \textit{et al} \cite{Sau} proposed that two-dimensional semiconductor quantum well can also support MFs, when both superconductivity and magnetism are induced through the proximity effect \cite{Alicea,Akhmerov}. Besides, Majorana edge modes are shown to emerge in one-dimensional (1D) semiconductor nanowire, which is deposited onto a SC and with a perpendicular magnetic field \cite{Oreg,Lutchyn}.

The above proposals to realize MFs in realistic materials are all based on superconductors. Therefore, it is interesting to ask the question: is it possible to realize the MFs without the help of any superconductivity?  A seminal work of Kitaev \cite{Kitaevb} demonstrated that MFs can emerge in a particular frustrated magnetic honeycomb model, indicating frustrated magnetic system may be a possible substitution when certain conditions are satisfied. In this work, we will give a more general answer to this question, by resorting to another type of quasi-particle excitations termed spinons, which only carry the spin rather than the charge degrees of freedom \cite{Wen}. Spinons are the basic excitations in the frustrated magnets or frustrated Mott insulators \cite{Leea}. Due to the frustration and the quantum fluctuation, the magnetic orders are melted, leading to non-symmetry-breaking quantum states. These quantum states, termed as the quantum spin liquid (QSL) \cite{Shimizu,Andersona,Andersonb,Yan}, remain disordered even at zero temperature, and the
elementary excitations they carry are the neutral spinful fermions (spinons). Spinons are very similar to electrons. They can enjoy a Fermi surface \cite{SLee,Motrunich} and
are generally accompanied by an emergent gauge field. Besides, similar to the Landau Fermi liquid, they may even develop instabilities after considering the interaction and the effect of the emergent gauge fluctuation. A typical instability is the formation of spinon pairs \cite{Lee,Galitski}, leading to the $\mathrm{Z}_2$ spin liquid state \cite{Wen,Galitski}. Experimentally, some candidate materials \cite{Shimizu,Okamoto,Helton,Itou} have been discovered, including 2D organic salts $\mathrm{EtMe_3Sb[Pd(dmit)_2]_2}$ and $\mathrm{\kappa-(BEDT-TTF)_2Cu_2(CN)_3}$, herbertsmithite $\mathrm{ZnCu_3(OH)_6Cl_2}$, as well as 3D hyperkagome lattice material $\mathrm{Na_4Ir_3O_8}$. Moreover, the spin-charge separation and the spinon excitation have been observed by the ARPES measurement in 1D $\mathrm{SrCuO_2}$ \cite{Jkim}.

In the present work, we explore the possibility to realize the MFs based on the spinon excitations, instead of the Bogoliubov quasi-particles in SC systems. It is revealed that, different from the conventional MFs which are combination of electrons and holes, another type of MFs can also be realized in the frustrated magnets, where a spinon and an anti-spinon form a MF state. Specifically, we consider a theoretical model where a QSL layer is in proximity to the surface of a 3D topological insulator (TI). In order to explore the topologically nontrivial state based on the spinons, we investigate the self-energy correction of the QSL due to the TI surface. It is found that the TI Dirac cone and its associated spin texture are transferred from the TI surface to the nearby QSL state. This effect, termed as the topological proximity effect (TPE), leads to the spinon-based MF (SMF) in the QSL layer. In contrary to the MF in the SC systems \cite{Fu,Read,Sau,Alicea,Akhmerov,Oreg,Lutchyn}, the SMFs are populated in the charge insulating QSL and are composite states of spinons and anti-spinons, so that they can play role in the transport of spinons, leading to unusual spin transport behavior. To demonstrate this point, we propose an experimental setup, where a two-peak spin conductance pattern is obtained. This serves as a clear experimental signature of the SMF.

The remaining part of this work is organized as following. In Sec.\uppercase\expandafter{\romannumeral2}, we introduce our theoretical model. In Sec.\uppercase\expandafter{\romannumeral3A}, a renormalization group (RG) analysis is performed in order to find out the most relevant order. Based on the RG result, a self-consistent mean-field study is presented in Sec.\uppercase\expandafter{\romannumeral3B}, where it is numerically found that the spin-momentum locking is inherited by the QSL state. To analytically account for this observation, in Sec.\uppercase\expandafter{\romannumeral3C} we derive an effective Hamiltonian of the dressed QSL state, taking into account the renormalization effect of the nearby TI surface. In Sec.\uppercase\expandafter{\romannumeral4}, the emergence of the SMF along the Zeeman boundary is proved. Furthermore, an experimental setup is put forward, where the calculated spin conductance is predicted to be a clear signature for the SMFs. Last, more discussion and conclusion are included in Sec.\uppercase\expandafter{\romannumeral5}.

\section{Model Hamiltonian}
To generate the Majorana fermions, three symmetries are usually broken, i.e., the $\mathrm{U(1)}$ gauge symmetry, the spin rotation symmetry and the time-reversal symmetry (TRS). Similarly, in order to realize the concept of SMF, we also need to break the three symmetries in QSL states, where the spinons act as the important quasi-particle expiations. All the known QSLs do not satisfy this condition. For example, the chiral spin liquid breaks the TRS while it respects the $\mathrm{U(1)}$ and spin rotation symmetry. Besides, the $Z_2$ spin liquid violates the $\mathrm{U(1)}$ but it preserves the TRS and spin rotation symmetry. However, considering that the Rashba interaction may further break the spin rotation symmetry, it indicates a model where a $Z_2$ spin liquid is in proximity to a topological insulator. If they couple to each other, one may expect a simultaneous breaking of the  $\mathrm{U(1)}$ gauge symmetry and the spin rotation symmetry. Then, the Majorana fermion may occur in a point defect or a domain wall where the TRS is locally violated.

Due to the above consideration, we now put forward a theoretical model, where a $Z_2$ QSL layer is in proximity to the surface of a strong topological insulator. We assume that, despite the possible interaction between them, the two subsystems are spatially separated so that measurement can be performed independently on each layer. The total Hamiltonian includes three parts, $H_{tot}=H_{TI}+H_{Z_2}+H_{c}$, where $H_{TI}$ describes the TI surface (lower layer)
\begin{equation}\label{eq1}
  H_{TI}=v_F\sum_{\mathbf{k}}c^{\dagger}_{\mathbf{k}}(z_2)\boldsymbol{\sigma}\cdot\mathbf{k}c_{\mathbf{k}}(z_2),
\end{equation}
where $c_{\mathbf{k}}(z_2)=[c_{\mathbf{k},\uparrow}(z_2),c_{\mathbf{k},\downarrow}(z_2)]^T$ is the two-component spinor of the Dirac cone in the surface. We use $z_2$ to represent the lower layer. $\boldsymbol{\sigma}$ denotes the true spin degree of freedom. We focus on the most interesting case where the Fermi energy lies at the Dirac point. Since Eq.\eqref{eq1} is the effective continuum model, a cutoff $\Lambda$ is implicit in the sum of $\mathbf{k}$.

The upper layer of the $Z_2$ spin liquid is described by the Hamiltonian $H_{Z_2}$, which can be derived from a theoretical model of the frustrated magnets $H_J$,
\begin{equation}\label{eq2}
H_{J}=\sum_{ij}J_{ij}\mathbf{S}_i(z_1)\cdot\mathbf{S}_j(z_1),
\end{equation}
where we use $z_1$ to represent the upper layer and the spin-half case is considered. Eq.\eqref{eq2} has been extensively studied \cite{faa,Hermele,Tay,Yan,Wenb}, and it is known that the strong frustration can give rise to the QSL state. Using the projected symmetry group approach \cite{Wen}, one can classify all the possible QSL states that are compatible with the symmetries of the Hamiltonian, among which the $Z_2$ spin liquid states are our main focus. Different types of $Z_2$ spin liquid states show different microscopic characteristic of the mean-field ansatz, which are related to the local fluctuation. In the long wavelength limit, the $Z_2$ spin liquids share common property such as the emergence of a spinon pairing term. Now we are going to derive an low-energy effective mean-field Hamiltonian for a typical $Z_2$ spin liquid \cite{Kli,Wen,Huh,Punk}.

First, we use the Schwinger fermion representation, $\mathbf{S}_i=f^{\dagger}_{i\alpha}\sigma_{\alpha\beta}f_{i\beta}$, with $f_{i,\uparrow/\downarrow}$ being the spinon operator, together with the Hilbert space constrict condition. $\sum_{\sigma}f^{\dagger}_{i\sigma}f_{i\sigma}=1$, which requires that there is only one spinon for any site $i$. The layer notation $z_1$ is not written for the purpose of brevity. Inserting this representation of spins, we obtain
\begin{equation}\label{eq3}
  H_J=\sum_{ij}-\frac{1}{2}J_{ij}(f^{\dagger}_{i\alpha}f_{j\alpha}f^{\dagger}_{j\beta}f_{i\beta}+\frac{1}{2}f^{\dagger}_{i\alpha}f_{i\alpha}f^{\dagger}_{j\beta}f_{j\beta}).
\end{equation}
Now let us introduce expectation values of $f^{\dagger}_{i\alpha}f_{j\beta}$ and $f_{i\alpha}f_{i\beta}$,
\begin{eqnarray}
  \zeta_{ij} &=& -2\epsilon_{\alpha\beta}\langle f_{i\alpha}f_{j\beta}\rangle, \\
  \chi_{ij} &=& 2\delta_{\alpha\beta}\langle f^{\dagger}_{i\alpha}f_{j\beta}\rangle,
\end{eqnarray}
where $\epsilon_{\alpha\beta}$ is the Levi-Civita symbol, and the sum of the repeated index are assumed. Then, the mean-field Hamiltonian of $H_J$ can be obtained, that reads
\begin{equation}\label{eq4}
\begin{split}
  H^{J}_{MF}&=\sum_{ij}-\frac{3}{8}J_{ij}[(\chi_{ji}f^{\dagger}_{i\alpha}f_{j\alpha}+\zeta_{ij}f^{\dagger}_{i\alpha}f^{\dagger}_{j\beta}\epsilon_{\alpha\beta}+H.C.)\\
  &-|\chi_{ij}|^2-|\zeta_{ij}|^2]+\sum_{i}\lambda^{\prime}(f^{\dagger}_{i\alpha}f_{i\alpha}-1),
\end{split}
\end{equation}
where $\lambda^{\prime}$ is the Lagrangian multiplier used to satisfy the Hilbert space constraints. The terms $|\chi_{ij}|^2$ and $|\zeta_{ij}|^2$ give us constants after the sum of sites. These constants will only play role when one tries to determine the mean-field parameters self-consistently. Here, in order to demonstrate the concept of spinon Majorana fermions and to be as general as possible, it is sufficient to treat the above Hamiltonian as a phenomenological model. Hence, we can safely neglect the constants $|\chi_{ij}|^2$ and $|\zeta_{ij}|^2$. Introducing a two-component spinor $\phi_i=[f_{i\uparrow},f^{\dagger}_{i\downarrow}]^T$ and the matrix
\begin{equation}\label{eq5}
U_{ij}=
\left(
  \begin{array}{cc}
    \chi^{\dagger}_{ij} & \zeta_{ij} \\
    \zeta^{\dagger}_{ij} & -\chi_{ij} \\
  \end{array}
\right),
\end{equation}
 the mean-field Hamiltonian can be further written into
\begin{equation}\label{eq6}
H^{J}_{MF}=\sum_{ij}-\frac{3}{8}J_{ij}(\phi^{\dagger}_{i}U_{ij}\phi_j+H.c.)+\sum_i\lambda^{\prime}\phi^{\dagger}_i\tau^3\phi_i.
\end{equation}
%where $\tau$ is the Pauli Matrix defined in Nambu space. $U_{ij}=(\tau^3+1)\chi^{\star}_{ij}/2+(\tau^3-1)\chi_{ij}/2+\tau^{+}\zeta_{ij}+\tau^{-}\zeta^{\star}_{ij}$, $\chi_{ij}=\langle f^{\dagger}_{i\alpha}(z_1)f_{j\alpha}(z_1)\rangle$, $\zeta_{ij}=\langle f_{i\alpha}(z_1)f_{j\beta}(z_1)\epsilon_{\alpha\beta}\rangle$ are the introduced RVB mean-field ansatz, where $\epsilon_{\alpha\beta}$ is the Levi-Civita tensor. $\lambda^{\prime}$ is the Lagrangian multiplier introduced to satisfy the Hilbert space constriction $\sum_{\sigma}f^{\dagger}_{i\sigma}(z_1)f_{i\sigma}(z_1)=1$ (for spin half system).
%
%Up to now, we has not discuss any detailed lattice and exchange interactions. From now on, we focus on
%the square lattice with the first nearest exchange interaction $J$ and the second nearest exchange interaction $J^{\prime}$. Because of $J^{\prime}\neq0$, frustration is present on this model. This model has been investigated by many literature. Among the results, predictions of $Z_2$ quantum spin liquid phase attracts our most attention.

In order to search for the quantum spin liquid ground state, Wen and the collaborators introduced the projected symmetry group (PSG) \cite{Wen}. Using this method, the first stable quantum spin liquid state is extracted from $H^J_{MF}$ in a square lattice (with the nearest neighbour interaction $J$ and the second nearest neighbour interaction $J^{\prime}$), termed as the T-P symmetric $Z_2$ mean-field ansatz \cite{Wenb}. Due to the simplicity of this ansatz, we use it as an example to derive the low-energy effective model (for other mean-field ansatz, similar low-energy effective model can be obtained). The T-P symmetric $Z_2$ mean-field ansatz reads,
\begin{eqnarray}
  U_{\mathbf{i},i+\hat{x}} &=& U_{\mathbf{i},\mathbf{i}+\hat{y}}=-\chi\tau^3, \\
  U_{\mathbf{i},\mathbf{i}+\hat{x}+\hat{y}} &=& \zeta\tau^1+\lambda \tau^2, \\
  U_{\mathbf{i},\mathbf{i}-\hat{x}+\hat{y}} &=& \zeta\tau^1-\lambda\tau^2.
\end{eqnarray}
which is invariant under the translation, parity and time-reversal transformation, together with the corresponding gauge transformations.
After inserting this mean-field ansatz into $H^{MF}_J$ and then making Fourier transformation to the momentum space, we arrive at
%\begin{eqnarray}
%% \nonumber to remove numbering (before each equation)
%  U_{\mathbf{i},i+\hat{x}} &=& U_{\mathbf{i},\mathbf{i}+\hat{y}}=-\chi\tau^3, \\
%  U_{\mathbf{i},\mathbf{i}+\hat{x}+\hat{y}} &=& \zeta\tau^1+\lambda \tau^2, \\
%  U_{\mathbf{i},\mathbf{i}-\hat{x}+\hat{y}} &=& \zeta\tau^1-\lambda\tau^2,
%\end{eqnarray}
\begin{equation}\label{eq7}
  H^J_{MF}=-\frac{3}{8}\sum_{\mathbf{k}}\phi^{\dagger}_{\mathbf{k}}(U_{\mathbf{k}}+\lambda^{\prime}\tau^3)\phi_{\mathbf{k}},
\end{equation}
where
\begin{equation}\label{eq8}
\begin{split}
  U(\mathbf{k})&=-2J\chi\tau^3\cos{k_x}-2J\chi\tau^3\cos{k_y}\\
  &+2J^{\prime}(\zeta\tau^1+\lambda\tau^2)\cos(k_x+k_y)\\
  &+2J^{\prime}(\zeta\tau^1-\lambda\tau^2)\cos(-k_x+k_y).
\end{split}
\end{equation}
After some algebra, the total Hamiltonian can be simplified into
\begin{equation}\label{eq9}
  H^{J}_{MF}=\sum_{\mathbf{k}}\phi^{\dagger}_{\mathbf{k}}
  \left(
  \begin{array}{cc}
  \epsilon_{\mathbf{k}} & \Delta_{\mathbf{k}}  \\
  \Delta^{\star}_{\mathbf{k}} & -\epsilon_{\mathbf{k}} \\
  \end{array}
  \right)\phi_{\mathbf{k}},
\end{equation}
where $\epsilon_{\mathbf{k}}$ and $\Delta_{\mathbf{k}}$ are
\begin{eqnarray*}
  \epsilon_{\mathbf{k}} &=& \frac{3}{4}J\chi(\cos k_x+\cos k_y)+\lambda^{\prime}, \\
  \Delta_{\mathbf{k}}&=& -\frac{3}{4}J^{\prime}\zeta[\cos(k_x+k_y)+\cos(-k_x+k_y)]\\
  &&+i\frac{3}{4}J^{\prime}\lambda[\cos(k_x+k_y)-\cos(-k_x+k_y)].
\end{eqnarray*}
%$\chi$, $\lambda$, $\zeta$, and $\lambda^{\prime}$ can be self-consistently obtained by minimizing the mean-field energy, which are dependent on $J$ and $J^{\prime}$. Here, our main focus in this work is not to obtain detailed QSL state, but to propose the spinon Majorana fermion (SMF) based on the QSL state. Hence, we do not perform the self-consistent calculation but treat the mean-field order parameters as phenomenological parameters.
Writing in terms of spinons, the Hamiltonian is reduced to the following continuum model of the $Z_2$ state in the long wave regime,
\begin{equation}\label{eq10}
 H_{Z_2}=\sum_{\mathbf{k}}[\mu f^{\dagger}_{\mathbf{k}\sigma}(z_1)f_{\mathbf{k}\sigma}(z_1)+\Delta f^{\dagger}_{\mathbf{k}\uparrow}(z_1)f^{\dagger}_{-\mathbf{k},\downarrow}(z_1)+H.C.,
\end{equation}
where $\mu=3J\chi/2+\lambda^{\prime}$ and $\Delta=-3J^{\prime}\zeta/2$, and the momentum cutoff $\Lambda$ is assumed. $f^{\dagger}_{\mathbf{k}\sigma}$, $f_{\mathbf{k}\sigma}$ are the creation and annihilation operators of the spinons with spin $\sigma$. We have recovered the layer notation $z_2$ for clarity. In the following, $\mu$, $\Delta$ are treated as phenomenological parameters, in order to be as general as possible.

%\begin{equation}\label{eq3}
%  H_m=\sum_{\mathbf{k}}[\epsilon_{\mathbf{k}} f^{\dagger}_{\mathbf{k}\sigma}(z_1)f_{\mathbf{k}\sigma}(z_1)+\Delta_{\mathbf{k}} f^{\dagger}_{\mathbf{k}\uparrow}(z_1)f^{\dagger}_{-\mathbf{k},\downarrow}(z_1)+H.C.],
%\end{equation}

%Eq.\eqref{eq11} can be derived from Eq.\eqref{eqn1} after the stable mean-field ansatz is inserted .
%Since our main aim is to study the coupling between the QSL layer and the TI surface layer where no momentum transfer takes place during the hopping process in the mean-field level (see Eq.\eqref{eqq3} below), and all the physical excitations of the TI Dirac cone are dominated by those near $\Gamma$ point in the low temperature regime, the spinons that are relevant in the coupling only originate from the excitations around the $\Gamma$ point. The effective Hamiltonian of the QSL in the long-wave limit then reads (see supplementary),
%For large region of parameters (see supplementary) of the QSL state, the energy minimum arises at the $\Gamma$ point, around which the dispersion is quite flat. This suggests that, in low temperature, excitations around $\Gamma$ point are most relevant. Therefore, in order to grab all the low-temperature physics, we can take the long-wave limit of the QSL Hamiltonian to obtain the continuum model of the QSL (see supplementary),

Eq.\eqref{eq10} describes spinons with a flat band and a spinon gap $\Delta$. This Hamiltonian is not oversimplified but captures all the essential physics in the low temperature regime. The dispersion of the QSL is negligible in the low-energy window since it is dominated by the linear energy spectrum of the TI electrons. As will be demonstrated later, even though being non-dispersive in the long wavelength limit, the QSL still plays a nontrivial role due to the spinon gap $\Delta$. In this sense, our following analysis does not rely on any short wavelength details of the $Z_2$ spin liquid, so that even a different $Z_2$ mean-field ansatz is discussed, our following calculation and result would not be qualitatively altered. Therefore, it is sufficient to study Eq.\eqref{eq10}, as a typical effective mean-field Hamiltonian of the $Z_2$ state.

Last, we should take into account the possible interaction between the electrons and the spinons in the two layers. Since spinons are chargeless, the Coulomb interaction is absent. The only possible interaction is the s-d coupling \cite{Chen} $H_c=g\sum_{\mathbf{r}}\mathbf{S}_c(\mathbf{r},z_1)\cdot\mathbf{S}(\mathbf{r},z_2)$, where the $\mathbf{S}_c(\mathbf{r},z_1)$ and $\mathbf{S}(\mathbf{r},z_2)$ are the spin density operators of the electrons and the spinons, respectively. $\mathbf{r}$ denotes the 2D real space coordinate parallel to the TI surface and the QSL layer. Using the Schwinger fermion representation \cite{Wen}, we arrive at the following s-d interaction,
%\begin{equation}\label{eq4}
%  H_c=g\sum_{\mathbf{r}}\mathbf{S}_c(\mathbf{r})\cdot\mathbf{S}_m(\mathbf{r}),
%\end{equation}
%where  is the spin density operator of the Dirac fermion. In terms of the pseudo fermion representation, we obtain the s-d interaction between electrons and spinons.
\begin{equation}\label{eq11}
  H_c=g\sum_{\mathbf{r}}c^{\dagger}_{\alpha}(\mathbf{r},z_2)c_{\beta}(\mathbf{r},z_2)f^{\dagger}_{\beta}(\mathbf{r},z_1)f_{\alpha}(\mathbf{r},z_1).
\end{equation}
where the Hilbert space constriction condition has been used. This interaction between electrons and spinons are quite different from the usual Hubbard-like Coulomb interaction between electrons. It includes the intra-layer spin-flip process which is absent in the latter.

%$c$-operators is defined in the TI surface while $f$-operators in the QSL layer,
%Since we consider the low energy behavior of the above models, a renormalization group (RG) analysis is useful. In order to do so, we introduce a momentum cutoff $\Gamma$. In the low energy limit, the momentum cutoff is recaled step by step in RG sense so that $\Gamma\rightarrow0$. So in the zero temperature case, all relevant momentum satisfy $k=k_0$, with $k_0\rightarrow0$. This condition leads to three  most relevant channels in the RG sense \cite{Shankar}, i.e., the forward scattering (FS), BCS, and the umklapp scattering (US) channel. Since in our model, only one Dirac cone is present in the TI surface so that the absent of the Fermi surface nesting excludes the instability in the umklapp scattering channel. Moreover, only attractive interactions develop observable orders in the BCS channel. So, the instability may only take place in the FS channel, which is the main focus in the following calculation. In the discussion, we will show that the Kondo-singlet state is the corresponding order that is developed in this channel.
Eqs.\eqref{eq1},\eqref{eq10} and \eqref{eq11} constitute the model we are going to investigate.  Ref.\cite{Fu} studies a similar setup where the TI surface is in proximity to a SC, and the pairing order is phenomenologically introduced into the TI surface. In comparison, our model contains two different types of basic excitations, which interact through the local s-d coupling $H_c$. To the best of our knowledge, the present model so far has not been studied in the literatures. Also, the effect of the s-d coupling $H_c$ between electrons and spinons needs to be carefully addressed. Moreover, instead of investigating the renormalization effect of the TI surface due to the QSL, it is more interesting to go the opposite way, i.e., studying how the TI surface would affect the QSL layer, since the topological nontrivial states based on the spinons are barely discussed.
\section{Topological proximity effect}
\subsection{Renormalization group theory of the leading instability}
To study the effect of the interaction Eq.\eqref{eq11}, we first perform a perturbative renormalization group analysis to find out the most relevant order preferred in the coupling system. Then, a self-consistent mean-field study on this specific order can be made.

Both Eq.\eqref{eq1} and Eq.\eqref{eq10} are effective continuum models, where a momentum cutoff $\Lambda$ is implicit. As what RG does, one can decrease $\Lambda$ step by step to obtain the effective action with renormalized parameters for the coarse-grained model, integrating out the degree of freedom with the larger momentum \cite{Shankar}.
%We first briefly comment on the RG method, which is the technique to find out the most relevant instability around a fixed point phase . For non-interacting case with $g=0$ in our model, the TI surface and the $Z_2$ spin liquid are well separated, and both are stable phases that act as the fixed point $S_0$ in our analysis. Now, we turn on $g$, and expand the interaction perturbatively around this fixed point $S_0$. By gradually integrating out the high energy degrees of freedom, we can obtain a scaling flow of $g$ with the scaling parameter $l$ (the energy scale decreases with the increase of $l$). If $g$ reduces with $l$, the fixed point $S_0$ will be robust against such interaction. However, when the flow of $g$ grows with $l$, it indicates that some instability will occur with the temperature lowered. To determine which type of instability would finally occur, we can further insert some possible vertices that describe certain orders. Then the scaling flow of such vertices can be calculated in the same way. The vertex that diverges the fastest (with the increase of $l$) shows us the most relevant instability. Through this method, we can determine which mean-field order is mostly favored in our model. In what follows, we show all the detailed calculation, which proves that the holon condensation is indeed the most relevant symmetry breaking phase.
Without interaction $g$, the model consists of $H_0=H_{TI}+H_{Z_2}$. $H_0$ describes a noninteracting fixed point. Resorting to the functional integral representation, the action $S_0$ of this fixed point reads as
\begin{equation}\label{eq12}
\begin{split}
  S_0&=\int d\tau\sum_{\mathbf{k}}v_F\overline{c}_{\mathbf{k}}(\partial_{\tau}+\boldsymbol{\sigma}\cdot\mathbf{k})c_{\mathbf{k}}\\
  &+\int d\tau\sum_{\mathbf{k}}\overline{f}_{\mathbf{k}}(\partial_{\tau}+\mu\tau^3+\Delta\tau^{+}+\Delta^{\star}\tau^-)f_{\mathbf{k}},
\end{split}
\end{equation}
with $c_{\mathbf{k}}=[c_{\mathbf{k},\uparrow},c_{\mathbf{k},\downarrow}]^T$ and $f_{\mathbf{k}}=[f_{\mathbf{k},\uparrow},f^{\dagger}_{-\mathbf{k},\downarrow}]^T$ being the Grassmann fields of electrons and spinons. $\boldsymbol{\tau}$ is the Pauli matrix denoting the Nambu space and $\tau^{\pm}=\tau^1\pm i\tau^2$. In the following, we omit the layer notations for the purpose of brevity. From $S_0$, the zeroth order Green's function of TI electrons and the spinons can be read off as,
\begin{eqnarray}
  \hat{G}^c(i\omega_n,\mathbf{k}) &=& 1/[i\omega_n-\boldsymbol{\sigma}\cdot\mathbf{k}], \\
  \hat{G}^f(i\omega_n,\mathbf{k}) &=& 1/[i\omega_n-\mu\tau^3-\Delta\tau^+-\Delta^{\star}\tau^-].
\end{eqnarray}

Now we turn on $g$.  The action for this coupling reads as,
\begin{equation}\label{eq13}
\begin{split}
  S_I&=g\int d\tau\sum_{\mathbf{k}_1,\mathbf{k}_2,\mathbf{k}_3} \\ &\times\overline{c}_{\alpha}(\mathbf{k}_1,\tau)c_{\beta}(\mathbf{k}_3,\tau)\overline{f}_{\beta}(\mathbf{k}_2,\tau)f_{\alpha}(\mathbf{k}_1+\mathbf{k}_2-\mathbf{k}_3,\tau),
\end{split}
\end{equation}
which can be represented by the Feynman diagram in Fig.1(a), where the solid lines and the dashed lines denote the propagators of the electrons and the spinons, respectively.

\begin{figure}[tbp]
\includegraphics[width=\linewidth]{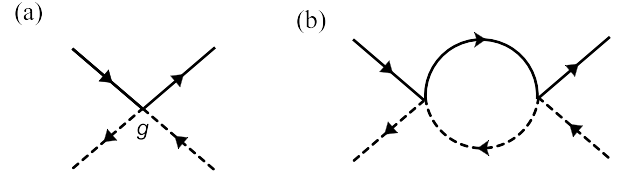}
\caption{(color online) (a) The Feynman diagram for the interaction between spinons and electrons. (b) The leading order renormalization of the interaction vertex.}
\end{figure}

Now we consider the renormalization of this interaction term. The lowest renormalization of $g$ comes from one-loop order, as is shown in Fig.1(b). Here, only the fast mode is integrated in the loop momentum.
To calculate the integral in Fig.1(b), we would encounter the the following term, $\sum_{\alpha,\beta}\langle G^c_{\alpha,\beta}(k)G^f_{\alpha,\beta}(k)\rangle$, where $\alpha,\beta$ are the spin notations (the momentum in the external legs are set to zero, which are irrelevant in RG sense \cite{Shankar}). The bracket $\langle...\rangle$ denotes the integration (sum) of the frequency $\omega_n$ and the fast mode momentum. Since the Green's function of the spinons is diagonal in the spin space, therefore we only need to calculate $\sum_{\alpha}\langle G^c_{\alpha,\alpha}(k)G^f_{\alpha,\alpha}(k)\rangle$. After inserting $\hat{G}^c$ and $\hat{G}^f$, and completing the integral, we arrive at
\begin{equation}\label{eq14}
  \langle G^c_{11}G^f_{11}\rangle=\langle G^c_{22}G^f_{22}\rangle=-\frac{1}{4\pi v_F^2}\frac{\Lambda^2}{\sqrt{\Delta^2+\mu^2}+\Lambda}dl,
\end{equation}
where $dl=d\Lambda/\Lambda$ is the RG flowing parameter.
Therefore, from the one-loop order renormalization, we know that $g$ is dressed to
\begin{equation}\label{eq15}
  \widetilde{g}=g+\frac{1}{2\pi v_F^2}\frac{\Lambda^2}{\sqrt{\Delta^2+\mu^2}+\Lambda}g^2dl.
\end{equation}
Now the renormalized total Hamiltonian can be written as
\begin{equation}\label{eq16}
\begin{split}
  H^r&=v_F\int d^2k c^{\dagger}_k\boldsymbol{\sigma}\cdot\mathbf{k}c_k\\
  &+\int d^2k(\mu f^{\dagger}_kf_k+\Delta f^{\dagger}_kf^{\dagger}_{-k}+\Delta^{\star}f_{-k}f_k)\\
  &+\widetilde{g}\int d^2k_1d^2k_2d^2k_3\\
  &\times c^{\dagger}_{\alpha}(k_1)c_{\beta}(k_3)f^{\dagger}_{\beta}(k_2)f_{\alpha}(k1+k2-k3).
\end{split}
\end{equation}
In the perturbative RG method, we still need to rescale the momentum and the fields back to the original scale \cite{Shankar}. This is achieved by setting $k^{\prime}=bk$, and $c^{\prime}=y^{-1}_1c, f^{\prime}=y^{-1}_2f$, with $b=e^{dl}$. The $S_0$ fixed point is required to remain intact after scaling, which leads to $y_1=b^{3/2}$ and $y_2=b$. Then, the final dressed coupling strength g becomes,
\begin{equation}\label{eq17}
  g^{\prime}=b^{-1}\widetilde{g}=g-gdl+\frac{1}{2\pi v_F^2}\frac{\Lambda^2}{\sqrt{\Delta^2+\mu^2}+\Lambda}g^2dl,
\end{equation}
so that the RG flow of $g$ is obtained as
\begin{equation}\label{eq18}
  \frac{dg}{dl}=-g+\frac{1}{2\pi v^2_F}\frac{\Lambda^2}{\Lambda+\sqrt{\Delta^2+\mu^2}}g^2.
\end{equation}
Two conclusions can be drawn from this flow equation. Firstly, the first term suppresses the growth of $g$, which is due to the linear dispersion of the Dirac electrons. Secondly, the second term is proportional to $g^2$, indicating a tendency for the divergence of $g$. Taking into account both the first term and the second term, we know that when the bare value of $g$ is large enough, the second term would dominate, so that the interaction will be renormalized to stronger and stronger value, suggesting an instability. This will be verified by the following mean-field study, where we find that when $g$ is larger than the threshold $g_c$, the system will develop the holon condensation order.

Since we have obtained the RG flow of $g$, one can further check which mean-field channel will be the mostly favored one \cite{Song,Rui}.  To do so, we introduce the possible ``mean-field vertices" describing the mean-field orders.  For example, we compare the following three orders, i.e., the holon condensation, the induced magnetic ordering (MO) in TI and the MO in QSL,
\begin{eqnarray}
  \text{holon condensation} &:& \eta_1\sum_{\mathbf{k}}f_{\mathbf{k},\alpha}c^{\dagger}_{\mathbf{k},\alpha}, \\
  \text{MO in TI} &:& \eta_2\sum_{\mathbf{k}}c^{\dagger}_{\mathbf{k},\alpha}\sigma^i_{\alpha\beta}c_{\mathbf{k},\beta}, \\
  \text{MO in QSL} &:& \eta_3\sum_{\mathbf{k}}f^{\dagger}_{\mathbf{k},\alpha}\sigma^i_{\alpha\beta}f_{\mathbf{k},\beta},
\end{eqnarray}
These terms can be represented by two-leg vertices, as shown in Fig.2(a), Fig.2(b), Fig.2(c), respectively.

\begin{figure}[tbp]
\includegraphics[width=3.2in]{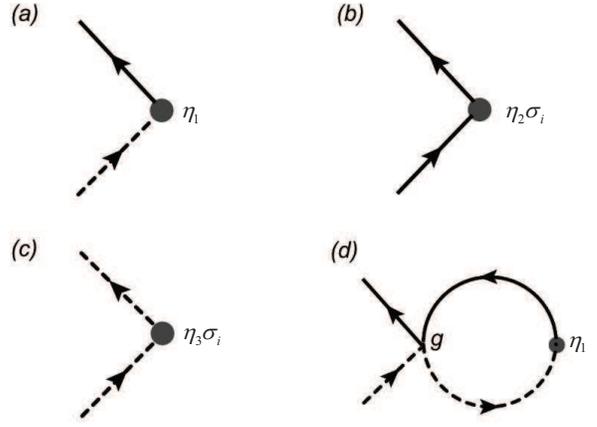}
\caption{(color online) (a), (b), (c) are the Feynman diagrams for the three introduced vertices. (d) the leading order renormalization of the holon condensation.}
\end{figure}

Now we calculate the flow of the order parameter $\eta_{i}$ ($i=1,2,3$) to one-loop order. The renormalization of $\eta_1$ is described by the above Fig.2(d). However, to one-loop order, $\eta_2$ and $\eta_3$ are not renormalized. Hence, compared to holon condensation order, the instability of MO can be neglected in the framework of perturbative RG.  After calculating the integral in the Feynmann diagram, the corresponding susceptibility \cite{Rui} $\Gamma_1$ of the holon condensation order can be obtained, whose flow with the RG parameter is shown in Fig.3.
%\begin{eqnarray}
%% \nonumber to remove numbering (before each equation)
%  \Gamma_1 &\propto& g, \\
%  \Gamma_2 &\propto& g^2, \\
%  \Gamma_3 &\propto& g^2.
%\end{eqnarray}
\begin{figure}[tbp]
\includegraphics[width=3.2in]{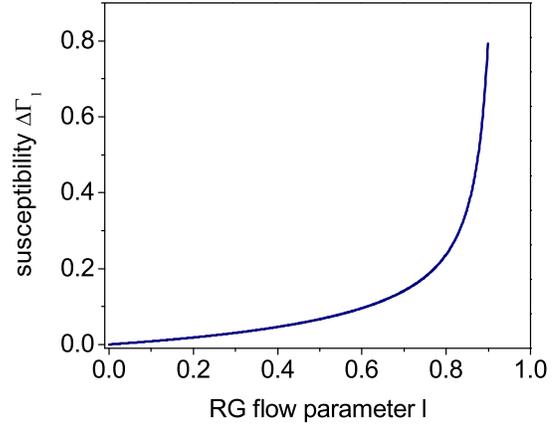}
\caption{(color online) The susceptibility of the holon condensation order versus the RG flow parameter. $\Delta\Gamma_1=\Gamma_1(l)-\Gamma_1(0)$. The parameters chosen are $g=0.158$, $\mu=0.135$ and $\Delta=0.04$ (in unit of the energy cutoff).}
\end{figure}
From the above susceptibility shown in Fig.3, we know that a divergence of $\eta_1$ would occur with the decreasing of the energy scale.

Through the above analysis, it is shown that the holon condensation channel is the most likely instability in our model. Based on this conclusion, a self-consistent mean-field investigation can be performed. Before proceeding, it is beneficial to discuss the physical meaning of the order in Eq.(26). Explicitly writing out $\eta_1$, we have $\eta_1=\sum_{\alpha}\langle c_{\alpha}(\mathbf{r},z_2)f^{\dagger}_{\alpha}(\mathbf{r},z_1)\rangle$. Obviously, Eq.(26) becomes an effective hopping term between an electron and a spinon in different layers. It describes the microscopic process where an electron is annihilated, a spinon is created, while at the same time a $\eta_1$ field is created, and vice versa. $\eta_1$ is the condensation of a spinon and a hole with the same spin, hence, it carries a net unit charge but no spin, physically equivalent to a holon. So, when an TI electron is hopping to a spinon in the QSL layer, a holon is created which remains in TI surface, and vice versa.

\subsection{Mean-field theory}
Based on the RG analysis, we decouple the s-d interaction in the holon condensation channel. After making the Hubbard-Stratonovich transformation of $H_c$, in the momentum space, we arrive at $H_c=-g\sum_{\mathbf{k}}[\eta_1 f_{\mathbf{k}\alpha}(z_1)c^{\dagger}_{\mathbf{k}\alpha}(z_2)+\eta^{\star}_1c_{\mathbf{k}\beta}(z_2)f^{\dagger}_{\mathbf{k}\beta}(z_1)]+g|\eta_1|^2$, where $\mathbf{k}$ is the 2D momentum.
\begin{figure}[tbp]
\includegraphics[width=3.4in]{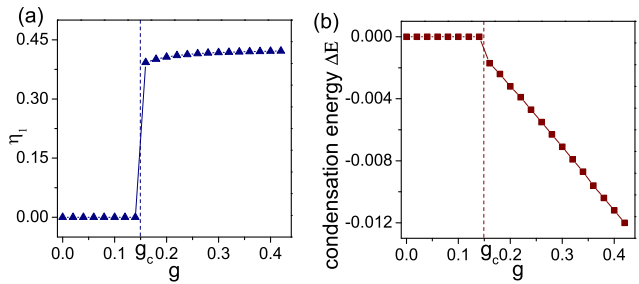}
\caption{(color online) (a) The self-consistent solution for the mean-field KH order parameter. (b) The condensation energy of the total TI and QSL system. The parameters chosen are $\Delta=0.4$, and $\mu=0$. For other values of $\Delta$ and $\mu$, qualitatively the same results are obtained, only with $g_c$ shifted accordingly. $v_F\Lambda$ is set as the unit.}
\end{figure}
Then, the mean-field Hamiltonian of the total system can be obtained. In the eight-dimensional hybridized basis $\Psi_{\mathbf{k}}=[\mathcal{C}_{\mathbf{k}}(z_2),\mathcal{F}_{\mathbf{k}}(z_1)]^T$, with $\mathcal{F}_{\mathbf{k}}(z_1)=[f_{\mathbf{k}\uparrow}(z_1),f_{\mathbf{k}\downarrow}(z_1),f^{\dagger}_{-\mathbf{k}\downarrow}(z_1),-f^{\dagger}_{-\mathbf{k}\uparrow}(z_1)]^T$ and $\mathcal{C}_{\mathbf{k}}(z_2)=[c_{\mathbf{k}\uparrow}(z_2),c_{\mathbf{k}\downarrow}(z_2),c^{\dagger}_{-\mathbf{k}\downarrow}(z_2),-c^{\dagger}_{-\mathbf{k}\uparrow}(z_2)]^T$, it reads
\begin{equation}\label{eq19}
\begin{split}
  H_{MF}&=\sum_{\mathbf{k}}\Psi^{\dagger}_{\mathbf{k}}[(\frac{s^3-s^0}{2})(\mu\tau^3+\Delta\tau^{1})+s^{+}\hat{t}^{\dagger}+s^{-}\hat{t}\\
  &+(\frac{s^3+s^0}{2})v_F\tau^3\boldsymbol{\sigma}\cdot\mathbf{k}]\Psi_{\mathbf{k}}+g|\eta_1|^2A,
\end{split}
\end{equation}
where $\boldsymbol{s}$ is the Pauli matrix denoting the ``fermion-spinon" pseudo spin and $\boldsymbol{\tau}$ is the one representing the Nambu space. $A$ is the area of the interface, and $\hat{t}=[(\tau^3+1)g\eta_1+(\tau^3-1)g\eta^{\star}_1]/2$ is the hopping matrix between the upper and lower layers. $\eta_1$ can be self-consistently determined by minimizing the ground state energy of $H_{MF}$. The results are shown in Fig.4(a). We found that with increasing $g$, $\eta_1$ undergoes a jump at a threshold value $g_c$. For $g<g_c$, $\eta_1=0$, which indicates that the small s-d coupling is irrelevant. This is due to the vanishing density of states of the Dirac cone. For $g>g_c$, $\eta_1\neq0$, so that the holon condensation order is formed, i.e. the lower layer electron and the upper layer spinons hybridize with each other. The above mean-field calculation is in agreement with the RG analysis, where it is found that the small bare value of $g$ is irrelevant while a divergence would occur with $g$ increased. To check whether this hybridized state is stable or not, we further calculate the condensation energy of the whole system. It is found in Fig.4(b) that, for $g<g_c$ the condensation energy remains at zero while it starts to decrease for $g>g_c$. The negative condensation energy justify the stability of the holon condensation order.

%The above mean-field results show that the KH order is formed between the surfaces of TI and QSL when the s-d coupling is large enough. This KH order induces an effective hopping between the electron and the spinon.
Now let us shift our attention to the band structure and the spin textures of the hybridized system. The energy spectrum  can be numerically obtained by diagonalizing $H_{MF}$. The result for $g=0$ and $g\eta_1=0.3$ is shown in Fig.5(a) and Fig.5(b), respectively. As shown in Fig.5(a), the TI Dirac cone and the QSL bands are completely decoupled with each other, while in Fig.5(b), the formation of the holon condensation order leads to four dispersive bands. Since only the long wave excitation dominates in the low-energy window, we focus on the dispersion around the $\Gamma$ point. Several conclusions are listed as following. First, we calculate the wave function distribution corresponding to the bands in the green shaded region in Fig.5(b). As shown in Fig.6(a), it is known that the lower (brown) bands in the shaded region mainly come from the TI surface, while the excitation in the higher (blue) bands primarily originates from the QSL layer.  Second, it is found that a gap is opened in the gapless TI Dirac cone. Last, we evaluated the expectation values of the spin operator of the spinons in the QSL layer. As is displayed in Fig.7(a), the spin of the spinons in the QSL layer is found to be locked to the momentum after the holon condensation order is formed (see also the schematic plot of the spin texture in Fig7.(b)). This spin texture is similar to that in the Dirac cone of the TI surface, and it breaks the spin rotation symmetry in the $Z_2$ spin liquid. Since the $\mathrm{U}(1)$ gauge symmetry is broken in the $Z_2$ spin liquid, we arrive at a state where the $\mathrm{U}(1)$ symmetry and the spin rotation symmetry are simultaneously broken, which perfectly satisfy the condition to realize the Majorana fermions, if the TRS is further locally broken by a defect or domain wall.

\begin{figure}[tbp]
\includegraphics[width=3.4in]{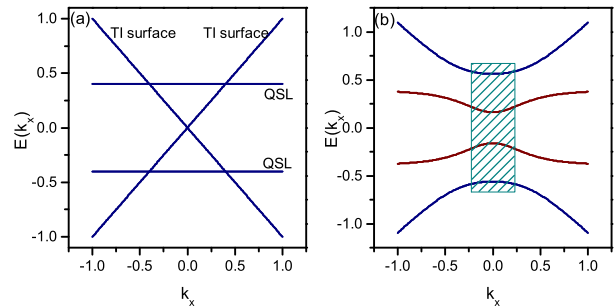}
\caption{(color online) (a), (b) are the calculated energy spectrum along $k_x$ ($k_y=0$) for $g=0$ and $g\eta_1=0.3$, respectively.  The other parameters chosen are $\mu=0$, $\Delta=0.4$.  For other values of parameters, qualitatively the same results are obtained.}
\end{figure}
\begin{figure}[tbp]
\includegraphics[width=3.4in]{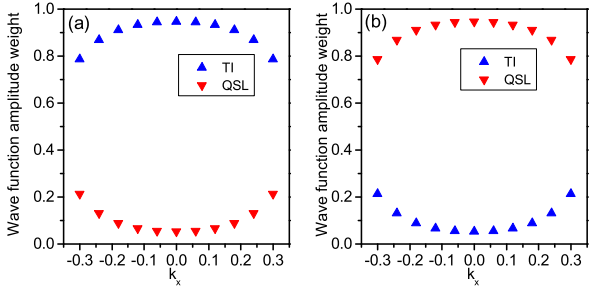}
\caption{(color online) The calculated contribution weight of QSL spinon and the TI electron to different bands. (a) The wave function weight corresponding to the lower brown bands in the green shaded region of Fig.5(b). (b) The wave function weight corresponding to the upper blue bands in the green shaded region of Fig.5(b).}
\end{figure}
\begin{figure}[tbp]
\includegraphics[width=3.4in]{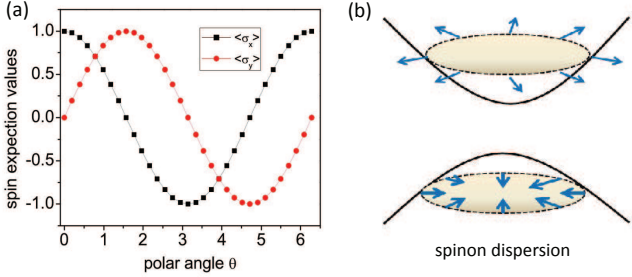}
\caption{(color online) (a) The calculated spin expectation values $\langle\sigma_x\rangle$ and $\langle\sigma_y\rangle$ associated to the blue bands in the green shaded region of Fig.5(b). The values are calculated along a circle around $\Gamma$ point. We use the polar coordinate with the radius $k=0.1$ and the polar angle $\theta$ is the x axis. (b) The schematic plot of the spin texture induced in the spinon dispersion.}
\end{figure}

Before proceeding, it is beneficial to discuss more about the role of the holon condensation order $\eta_1$ in terms of the spontaneous symmetry breaking. Before the mean-field order develops nonzero value, we can check that $H_{TI}$ breaks the spin rotation while preserves the $\mathrm{U}(1)$ gauge symmetry of the electron fields, and the $H_{Z_2}$ violates the $\mathrm{U}(1)$ gauge symmetry but respects the spin rotation symmetry of the spinon fields. After the mean-field order sets in, the system goes through a phase transition and a term occurs in the mean-field Hamiltonian as $H_c=-g\sum_{\mathbf{k}}[\eta_1 f_{\mathbf{k}\alpha}(z_1)c^{\dagger}_{\mathbf{k}\alpha}(z_2)+\eta^{\star}_1c_{\mathbf{k}\beta}(z_2)f^{\dagger}_{\mathbf{k}\beta}(z_1)]+g|\eta_1|^2$, . This term is a result of the spontaneous symmetry breaking as one can find that it further breaks the $\mathrm{U}(1)$ gauge symmetry of the electron fields and also the spin rotation symmetry of the spinon fields. Hence, the two symmetries are now both violated in the whole system due to the nonzero order parameter $\eta_1$, resulting in a satisfactory platform to generate Majorana excitations.

\subsection{Effective Hamiltonian for the dressed state}
The above numerical results indicate that the spinon seems to inherit the spin-momentum locking from the TI surface. To account for such numerical observation, we now shift our attention to analytically derive an effective Hamiltonian of the spinons near $k=0$.  In order to do so, we use the functional path integral method and integrate out the renormalized TI electrons (with a gap opened up). Moreover, we numerically verified that all the results are qualitatively the same for both $\mu=0$ and $\mu\neq0$. Hence, we analytically derive the $\mu=0$ case, for the purpose of brevity.

From the mean-field Hamiltonian of the total system Eq.\eqref{eq19}, we can obtain the functional representation in terms of the following basis, $\mathcal{F}_{\mathbf{k}}(z_1)=[f_{\mathbf{k}\uparrow}(z_1),f_{\mathbf{k}\downarrow}(z_1),f^{\dagger}_{-\mathbf{k}\downarrow}(z_1),-f^{\dagger}_{-\mathbf{k}\uparrow}(z_1)]^T$ and $\mathcal{C}_{\mathbf{k}}(z_2)=[c_{\mathbf{k}\uparrow}(z_2),c_{\mathbf{k}\downarrow}(z_2),c^{\dagger}_{-\mathbf{k}\downarrow}(z_2),-c^{\dagger}_{-\mathbf{k}\uparrow}(z_2)]^T$. For the purpose of brevity, we neglect the notation $z_1$ and $z_2$ in the following. The imaginary-time action of the total system then reads,
\begin{equation}\label{eq20}
\begin{split}
  S&=\sum_{\mathbf{k}}\int d\tau\overline{\mathcal{C}}_{\mathbf{k}}(\partial_{\tau}+v_F\tau^3\boldsymbol{\sigma}\cdot\mathbf{k})\mathcal{C}_{\mathbf{k}}\\
  &+\sum_{\mathbf{k}}\int d\tau\overline{\mathcal{F}}_{\mathbf{k}}(\partial_{\tau}+\Delta\tau^1)\sigma^0\mathcal{F}_{\mathbf{k}}\\
  &+\sum_{\mathbf{k}}\int d\tau[\overline{\mathcal{C}}_{\mathbf{k}}\hat{t}^{\dagger}\mathcal{F}_{\mathbf{k}}+\overline{\mathcal{F}}_{\mathbf{k}}\hat{t}\mathcal{C}_{\mathbf{k}}].
\end{split}
\end{equation}
First, we consider the renormalization effect of the TI electrons due to the QSL spinons by integrating out the QSL spinons.  From the last two lines in Eq.\eqref{eq20}, an effective free energy correction can be obtained after integrating out the Grassmann fields $\overline{\mathcal{F}}_{\mathbf{k}}$ and $\mathcal{F}_{\mathbf{k}}$, which reads as,
\begin{equation}\label{eq21}
\begin{split}
  \Delta F&=g^2|\eta_1|^2T\\
  &\times\sum_{\mathbf{k}}\overline{C}_{\mathbf{k}}[\sum_{n}\frac{i\omega_n-\Delta e^{-i2\alpha}\tau^{+}-\Delta e^{i2\alpha}\tau^{-}}{(i\omega_n)^2-(\Delta^2+\mu^2)}]\mathcal{C}_{\mathbf{k}}
\end{split}
\end{equation}
where $\alpha$ is the phase of $\eta_1$. Completing the sum of the Matsubara frequency, and taking the zero temperature limit, the above free energy is obtained as
\begin{equation}\label{eq22}
\begin{split}
  \Delta F&=\frac{1}{2}g^2|\eta_1|^2\sum_{\mathbf{k}}\overline{\mathcal{C}}_{\mathbf{k}}[e^{-i2\alpha}\tau^{+}+e^{i2\alpha}\tau^{-}]\mathcal{C}_{\mathbf{k}}\\
  &+\frac{1}{2}g^2|\eta_1|^2\sum_{\mathbf{k}}\overline{\mathcal{C}}_{\mathbf{k}}\mathcal{C}_{\mathbf{k}}.
\end{split}
\end{equation}
The second term can be neglected. This is because, after expanding in terms of the components, it leads to the sums of   $\overline{c}_{\mathbf{k}\uparrow}c_{\mathbf{k}\uparrow}+c_{\mathbf{k}\uparrow}\overline{c}_{\mathbf{k}\uparrow}$, and $\overline{c}_{\mathbf{k}\downarrow}c_{\mathbf{k}\downarrow}+c_{\mathbf{k}\downarrow}\overline{c}_{\mathbf{k}\downarrow}$, which, due to the anticommutative algebra of the Grassmann fields, vanish. Now, taking into account the above free energy correction due to the QSL spinon, we arrive at the renormalized Hamiltonian of the TI electrons,
\begin{equation}\label{eq23}
  H^r_{TI}=\sum_{\mathbf{k}}\mathcal{C}^{\dagger}_{\mathbf{k}}[v_F\tau^3\boldsymbol{\sigma}\cdot\mathbf{k}+\frac{1}{2}g^2|\eta_1|^2(e^{-i2\alpha}\tau^{+}+e^{i2\alpha}\tau^{-})]\mathcal{C}_{\mathbf{k}}.
\end{equation}
The second term of the above Hamiltonian brings a gap to the original gapless Dirac cone in the TI surface. This accounts for the observation of a gap in the renormalized TI dispersion around $\Gamma$ point. From Eq.\eqref{eq23}, it is known that the $\tau^{+}$ and $\tau^{-}$ term breaks the $\mathrm{U}(1)$ symmetry of the TI electrons. Moreover, the spin rotation symmetry is broken by the first term. Hence, as has been proved by Ref.\cite{Fu,Read}, Majorana fermions can be generated at point defect or domain wall, where the TRS is broken. The Majorana fermions observed in the TI side can be formally regarded as the superposition of electron and hole states in the TI surface, which are not our main focus in this work. In the following, we are going to show that MFs can also be generated in the QSL side, which are superposition of spinon and anti-spinon states in the QSL layer.

As the second step, we now consider the self-energy correction of the QSL spinons due to the TI electrons. Since the renormalized TI electrons will in turn renormalize the QSL spinons through the effective hopping process, one need to integrate out the Grassmann fields $\overline{\mathcal{C}}_{\mathbf{k}}$ and $\mathcal{C}_{\mathbf{k}}$ in the following action in order to obtain a free energy correction.
\begin{equation}\label{eq24}
\begin{split}
  S^{r}&=\sum_{\mathbf{k}}\int d\tau\overline{\mathcal{C}}_{\mathbf{k}}[\partial_{\tau}+\frac{1}{2}g^2|\eta_1|^2(e^{-i2\alpha}\tau^{+}+e^{i2\alpha}\tau^{-})\\
  &+v_F\tau^3\boldsymbol{\sigma}\cdot\mathbf{k}]\mathcal{C}_{\mathbf{k}}+\sum_{\mathbf{k}}\int d\tau[\overline{\mathcal{C}}_{\mathbf{k}}\hat{t}^{\dagger}\mathcal{F}_{\mathbf{k}}+\overline{\mathcal{F}}_{\mathbf{k}}\hat{t}\mathcal{C}_{\mathbf{k}}].
\end{split}
\end{equation}
Integrating out electron fields, we arrive at the free energy correction as
\begin{equation}\label{eq25}
  \Delta F^{\prime}=g^2|\eta_1|^2T\sum_{\mathbf{k}}\overline{\mathcal{F}}_{\mathbf{k}}[\sum_n\frac{i\omega_{n}+v_F\tau^3\boldsymbol{\sigma}\cdot\mathbf{k}+|\Delta_{e}|\tau^1}{(i\omega_n)^2-(|\Delta_{e}|^2+v_F^2k^2)}]\mathcal{F}_{\mathbf{k}},
\end{equation}
where $|\Delta_{e}|=\frac{1}{2}g^2|\eta_1|^2$. After completing the sum of the Matsubara frequency, and taking the zero temperature limit, the above free energy is obtained as
\begin{equation}\label{eq26}
\begin{split}
  \Delta F^{\prime}&=-\frac{1}{2}g^2|\eta_1|^2\sum_{\mathbf{k}}\overline{\mathcal{F}}_{\mathbf{k}}\frac{v_F\tau^3\boldsymbol{\sigma}\cdot\mathbf{k}+|\Delta_e|\tau^1}{\sqrt{|\Delta_e|^2+v^2_Fk^2}}\mathcal{F}_{\mathbf{k}}\\
  &+\frac{1}{2}g^2|\eta_1|^2\sum_{\mathbf{k}}\overline{\mathcal{F}}_{\mathbf{k}}\mathcal{F}_{\mathbf{k}},
\end{split}
\end{equation}
where the second term is again neglected due to the Grassmann algebra (after expansion). For $g=0$, $\Delta F^{\prime}=0$ so that all the above renormalization effect vanish, accounting for the decoupled case in Fig.5(a). For $g\neq0$ and $\eta_1\neq0$, we can already find from the above equation a Rashba-like term. This term is due to the fact that TI electrons renormalize the QSL spinons. Since all the most relevant excitations reside around the $\Gamma$ point, we can write down the free energy correction in the long wave approximation,
\begin{equation}\label{eq27}
  \Delta F^{\prime}=\sum_{\mathbf{k}}\overline{\mathcal{F}}_{\mathbf{k}}[v_F\tau^3\boldsymbol{\sigma}\cdot\mathbf{k}+\frac{1}{2}g^2|\eta_1|^2\tau^1]\mathcal{F}_{\mathbf{k}}.
\end{equation}
Finally, taking into account the above correction, the renormalized spinons in the QSL is described by the effective Hamiltonian,
\begin{equation}\label{eq28}
H_{eff}=\sum_{\mathbf{k}}\mathcal{F}^{\dagger}_{\mathbf{k}}[v_F\tau^3\boldsymbol{\sigma}\cdot\mathbf{k}+(\Delta+\frac{1}{2}g^2|\eta_1|^2)\tau^1]\mathcal{F}_{\mathbf{k}}.
\end{equation}
The above effective Hamiltonian shows that there exists a topological proximity effect between the TI surface and the QSL layer.  Besides, it is worth noting that, for $g=0$ or $\eta_1=0$ the first term in Eq.\eqref{eq28} is actually zero, since the free energy correction Eq.\eqref{eq26} is zero. This is not reflected in Eq.\eqref{eq28}, because the long wave approximation does not hold for $g=0$. Hence, Eq.\eqref{eq28}, as a result in the long-wave limit, is only correct for the hybridized phase where $g$ and $\eta_1$ are nonzero.

Several conclusions can be drawn from Eq.\eqref{eq28}. First, the original spinon gap is found to be renormalized by the amount $g^2|\eta_1|^2/2$. Second, the first term in Eq.\eqref{eq28}, as the self-energy correction due to the nearby TI surface, describes a Dirac cone term in the spinon dispersion. This Rashba-like term clearly shows the emergence of the TPE, i.e., the spin-momentum locked Dirac cone of the TI surface is passed on to the QSL layer, and accounts for the extracted spin texture of the spinons in Fig.7. Interestingly, a similar effect has been experimentally observed in the junction between a topological insulator $\mathrm{TlBiSe}_2$ and a metal $\mathrm{Bi}$ layer \cite{Shoman}. This observation of TPE makes our proposal feasible and timely in experiment.
%  Also, the above theoretical analysis can serve as a satisfactory explanation of the experiment in Ref.\cite{Shoman}, with a straight forward extension to replace the QSL by the  $\mathrm{Bi}$ layer.

\section{Spinon Majorana fermion and its spin transport}
From Eq.\eqref{eq28}, the spin texture and the spin-momentum locking are introduced to the spinons in QSL due to TPE. This is nontrivial in the sense that it breaks spin rotation symmetry in a $Z_2$ state where $\mathrm{U}(1)$ symmetry is absent, so that Eq.\eqref{eq28} serves as a satisfactory platform to generate Majorana excitations. As a straigtforward generalization of Ref.\cite{Fu,Read}, one can prove from Eq.\eqref{eq28} that a chiral Majorana mode (CMM) will take place if a boundary is considered, outside which there is an Zeeman field. The Zeeman boundary locally breaks the TRS in the hybridized state, where both the $\mathrm{U}(1)$ and the spin rotation symmetry are simultaneously violated. Hence, all the symmetry requirement of the Majorana fermions are satisfied. In this work, we mainly discuss the CMM along the Zeeman boundary, as a realization of the gapless chiral SMFs.
%If one denote $h$ So, Eq.\eqref{eq5} shows that the coupling between TI and the Mott insulator is equivalent to the process where an electron
%Here, the spin-momentum locking leads to a Dirac-type eigenequation, which may have zero energy soliton solutions. This topic has been discussed in many previous works. Ref.\cite{Fu} studies the Majorana fermions in superconductor system. Ref.\cite{Hou} discusses graphene with a chirality-mixing texture in the background, and Ref.\cite{Seradjeh} investigates the excitonic condensation between two opposite TI surfaces. Despite the similar mathematics, the physics in our work is different. In contrary to Ref.\cite{Hou,Seradjeh}, since our theory lies in Nambu space, no charge fractionalization but Majorana excitation will occur. Besides, different from Ref.\cite{Fu} whose Majorana excitation is the composition of an electron and a hole, the Majorana mode here would be the  composite state of a spinon and an anti-spinon.

To understand the excitations in the CMM, we now discuss the Majorana bound state by formally assuming a vortex core in the spinon pairing order parameter, i.e., $\Delta=\Delta_0e^{i\phi}$. After solving the zero energy BdG equation $\mathcal{H}(\mathbf{r})\psi(\mathbf{r})=0$ related to Eq.\eqref{eq28}, we arrive at the bound state solution
\begin{eqnarray}
  \psi_{1}(r,\phi) &=& Ae^{-\Delta^{\prime}r/v_F}\gamma_{1} \\
  \psi_{2}(r,\phi) &=& -Ae^{i\phi}e^{-\Delta^{\prime}r/v_F}\gamma_2,
\end{eqnarray}
with $A$ being the normalization factor, $\Delta^{\prime}=\Delta+g^2|\eta_1|^2/2$, and $\gamma_1=(0,i,1,0)^T$, $\gamma_2=(1,0,0,-i)^T$. Through a $\mathrm{U}(1)$ phase transformation, one can obtain the $\gamma$ operators in the second quantized form, i.e.,
\begin{equation}\label{eq29}
  \gamma_{1,2}=cf^{\dagger}_{\downarrow,\uparrow}+c^{\star}f_{\downarrow,\uparrow},
\end{equation}
where $c$ is an arbitrary coefficient. It is obvious that $\gamma_{1,2}^{\dagger}=\gamma_{1,2}$. Therefore, this bound state is a Majorana excitation. Besides, different from the conventional realization of the MF in SC systems, Eq.\eqref{eq29} shows that the MF predicted here is the equal-weight combination of a spinon and an anti-spinon, rather than an electron and a hole. Hence, we term this Majorana fermion as the SMF.

Being a self-conjugate quasi-particle, SMF enjoys similar physical properties with the conventional MF. However, since SMFs are populated in the charge insulating QSL state and are composite states of spinons and anti-spinons, they can play important role in the transport of spinons, leading to unusual spin transport.  As shown in Fig.8(a), we consider the junction between 1D quantum magnet material (such as the 1D $\mathrm{LiCuSbO_4}$ \cite{Dutton}) and the chiral spinon Majorana mode (CSMM).  In Fig.8(a),  a round QSL region is placed in proximity to the TI substrate, outside which there is a Zeeman field. Around the boundary, the CSMM will emerge  \cite{Read}. It is connected to the left and right 1D $\mathrm{LiCuSbO_4}$, which are well separated from the TI surface, so that the effect of the TI surface can be neglected in the transport process. The 1D quantum magnets are further coupled to the left and right metal leads and each lead couples to two ferromagnetic electrodes. A current source is used between the two left FM electrodes, which can create a spin-resolved chemical potential \cite{Chen} $\nu_{\alpha,\sigma}$, with $\alpha=L,R$ and $\sigma=\uparrow,\downarrow$ in the 1D $\mathrm{LiCuSbO_4}$.

The Hamiltonian contains three parts, $H=H_0+H_{\gamma}+H_T$, where $H_0$, $H_{\gamma}$ and $H_T$ are the Hamiltonians of the one-dimensional quantum magnets, the CSMM, and the coupling between them at the left and right contacting point $r_0$ (a) and $r_1$ (b), respectively.
\begin{eqnarray}
  H_{0} &=& \sum_{\alpha,\mathbf{k},\sigma}(\xi_{\mathbf{k}}-\nu_{\alpha,\sigma})f^{\dagger}_{\alpha k\sigma}f_{\alpha k\sigma}, \\
  H_{\gamma} &=& iv\int^l_0dr\gamma(r)\partial_r\gamma(r), \\
  H_T &=& t\sum_{\sigma}[f^{\dagger}_{L\sigma}(r_0)\gamma(a)+f^{\dagger}_{R\sigma}(r_1)\gamma(b)]+H.C..
\end{eqnarray}
%We assume the presence of a spin-resolved chemical potential $\nu_{\alpha,\sigma}$, with $\alpha=L,R$ and $\sigma=\uparrow,\downarrow$, which can be induced by the scheme introduced in Ref.\cite{Chen}.
The spin current  $I_S$ can be calculated as $I_S=-<dS^z_{L,tot}/dt>$, where $S^z_{L,tot}=(1/2)\sum_{\mathbf{k}}[f^{\dagger}_{L\uparrow}(\mathbf{k})f_{L\uparrow}(\mathbf{k})-f^{\dagger}_{L\downarrow}(\mathbf{k})f_{L\downarrow}(\mathbf{k})]$ is the total spin in the left lead. Then the Heisenberg motion of equation leads to
\begin{equation}\label{eq30}
  I_S=\frac{t}{2}Re[G^{<}_{L\uparrow}(r_0,0;a,0)-G^{<}_{L\downarrow}(r_0,0;a,0)],
\end{equation}
where $G^{<}_{L\uparrow,\downarrow}(x_1,t_1;x_2,t_2)=i<f^{\dagger}_{L\uparrow,\downarrow}(x_1,t_1)\gamma(x_2,t_2)>$ is the lesser Green's function. In order to obtain $G^{<}_{L\uparrow,\downarrow}(x_1,t_1;x_2,t_2)$, we resort to the non-equilibrium Green's function method by extending the real time $t$ to contour time $\tau$. Using the Keldysh equation and the analytic theorem \cite{Haug}, one is able to express $G^{<}_{L\uparrow,\downarrow}(x_1,t_1;x_2,t_2)$ by the Green's function of the quantum magnets and the Majorana mode, which finally gives us the spin conductance through the CSMM.

%without any loss of generality, we consider the case where $\nu_{L,\uparrow}-\nu_{R,\uparrow}=\Delta\nu$ and $\nu_{L,\downarrow}-\nu_{R,\downarrow}=\Delta\nu+V$ with $V$ being the induced spin bias in the left 1D quantum magnets.
Fig.8(b) shows the spin conductance versus the chemical potential difference between the left and right 1D $\mathrm{LiCuSbO_4}$, $\Delta\nu$, where it is found that, similar to the charge conductance through the CMM \cite{Law}, the CSMM generates periodical spin conductance peaks. In each complete period, two opposite peaks are obtained (the distance between the two peaks depends on the value of $\nu_{L\uparrow}-\nu_{L\downarrow}$), which can serve as a clear experiment signature of the SMF. Finally, the spin conductance can be detected by the voltmeter that is connected to the right two ferromagnetic electrodes \cite{Chen,Tinkham}.
\begin{figure}[tbp]
\includegraphics[width=\linewidth]{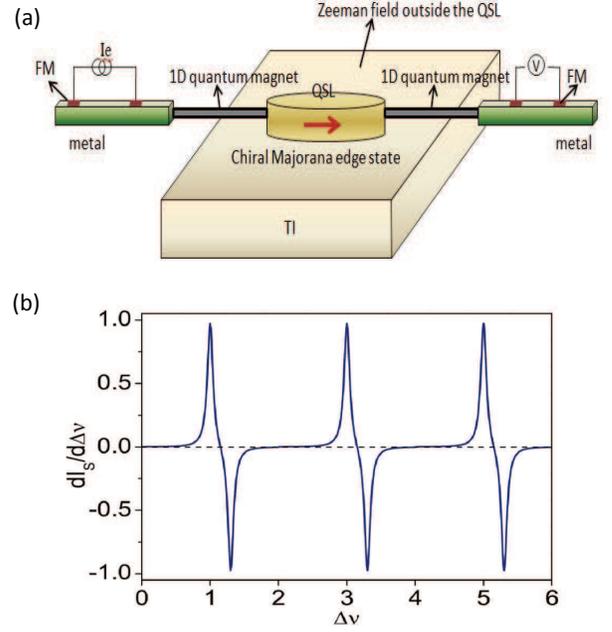}
\caption{(color online) (a) The schematic plot of the junction setup between 1D quantum magnets and the chiral spinon Majorana mode. (b) The spin conductance $dI_S/d\Delta\nu$ versus the chemical potential difference $\Delta\nu$. $dI_S/d\Delta\nu$ is in unit of $\hbar/2$ and $\Delta\nu$ is in unit of $\pi\hbar v/l$.}
\end{figure}

\section{Conclusion and discussion}
Further discussion is in what follows. First, immediate generalization can be made to show that the TPE would also occur in the interface between the TI and a normal metal. Hence, the TPE predicted here is not an exclusive effect between TI and QSL but is quite universal. Second, an advantage of our proposal is that we do not require high purity of the underlying compound. The Dirac electrons in the TI surface are known to be free from backscattering and immune to non-magnetic disorder. Besides, using the analogy of the disorder in SC case and recalling the Anderson theorem \cite{Balatsky}, one can expect the $Z_2$ spin liquid would show robustness to disorder. Third, an estimation of the spinon gap renormalization can be made. For $g=0.2$, the gap shift $g^2|\eta_1|^2/2$ is estimated to be around 0.032meV. Taking into account the original gap in possible $Z_2$ spin liquid candidate \cite{yqi}, we expect the in-gap physics can be observed below $T=1$K. Last, from Eq.\eqref{eq23}, it is known that a gap is opened up in the renormalized TI Dirac electrons, which is similar to the topological superconductor scenario \cite{Fu}. Hence, despite the SMF in the QSL layer, MF can also be generated in the TI surface.

To summarize, as an analogy of the MF in SC systems, we theoretically put forward the concept of SMF in QSL. The experimental setups to realize and detect the SMF are proposed. A TPE is predicted, where we show that the TI can induce a Rashba-like term into the nearby material around $\Gamma$ point. Therefore, when a QSL layer is in proximity to TI, the induced spin-momentum locking generates a topologically-nontrivial state based on spinons, where SMF emerges as the quasi-particle excitation along the Zeeman boundary. As a demonstration of the experimental detection, we show that the SMF plays a major role in the transport of spinons, leading to distinguished peaks in the spin conductance. Last, since the SMF is realized in frustrated magnetic materials rather than SCs, it may open a different route in the field of the topological quantum computation \cite{Nayak}.
%  de to realize the Majorana fermion in the frustrated magnets through the TPE. The predicted MF can be regarded as a combination of a spinon and an anti-spinon, hence is termed as SMF.  The SMF and the conventional MF manifest themselves in different transport properties. Since the SMF is the topological excitation in QSL state (Eq.\eqref{eq9}), which is charge insulating, the SMF is exptected to be . However, as discussed before, it can . These properties are absent in the convention MFs, which can take part in the transport of electrons but not spinons.
%As shown in Fig.4(b), the opposite-two-peak signature arises due to the spin-resolved chemical potential. When $V\neq0$, the two spin sectors are shifted along $\delta\nu$ axis. Since the up and down sectors contribute the positive and negative conductance respectively, the shift leads to the two-opposite-peak pattern. This pattern serves as the clear signatures of the proposed SMF

\begin{acknowledgments}
Rui Wang acknowledges L. Hao, Peter P. Orth, O. Erten, J.-X. Zhu, H. Li, and Y. Zhao for fruitful discussion. This work is supported by the Texas Center for Superconductivity at the University of Houston and the Robert A. Welch Foundation (Grant No. E-1146), the National Natural Science Foundation of China (Grants No. 60825402, No. 11574217, No. 11574108), the Natural Science Foundation of Anhui Province (Grant No. 1408085QA12), and the Natural Science Research Project of Higher Education Institutions of Anhui Province (Grant No. KJ2015A120).

% This work was supported by 973 Program under Grant No. 2011CB922103, and by
%NSFC (Grants No. 60825402, No. 11023002, No. 91021003 and No. 11574108).
\end{acknowledgments}


%Figure Captions \newline

\end{document}